\documentstyle[psfig]{mn}

\def\eg{{e.g.\thinspace}}

\def\ie{{i.e.\thinspace}}

\def\mJy{\,\hbox{mJy}}
\def\hMpc{\,h^{-1}\,\hbox{Mpc}}
\def\hMsol{h^{-1}\,\hbox{M}_{\odot}}
\def\kmsMpc{\,\hbox{km}\,\hbox{s}^{-1}\,\hbox{Mpc}^{-1}}

\def\harc2{\, h \, {\rm arcmin}^2}
\def\arc2{{\rm arcmin}^2}

\begin{document}
\journal{preprint astro-ph/0102352}
\title[SZ Predictions for the {\it Planck} satellite]
      {Sunyaev--Zel'dovich Predictions for the {\it Planck Surveyor} 
	Satellite using the Hubble Volume Simulations}
\author[S.~T.~Kay, A. R.~Liddle and P.~A.~Thomas]{Scott
	T.~Kay,\thanks{email: S.T.Kay@sussex.ac.uk} Andrew R.~Liddle
	and Peter A.~Thomas\\ Astronomy Centre, University of Sussex,
	Falmer, Brighton BN1 9QJ\\}
\date{\today}
\maketitle
\begin{abstract}
We use the billion-particle Hubble Volume simulations to make
statistical predictions for the distribution of galaxy clusters 
that will be observed by the {\it Planck Surveyor} satellite 
through their effect on the cosmic microwave background ---
the Sunyaev--Zel'dovich effect. We utilize the lightcone datasets
for both critical density ($\tau$CDM) and flat low-density ($\Lambda$CDM)
cosmologies: a `full-sky' survey out to $z \sim 0.5$, two
`octant' datasets out to beyond $z=1$ and a 100 square degree dataset 
extending to $z \sim 4$. Making simple, but robust, assumptions regarding
both the thermodynamic state of the gas and the detection of objects
against an unresolved background, we present the expected number of 
SZ sources as a function of redshift and angular size, and also by flux 
(for both the thermal and kinetic effects) for 3 of the relevant HFI frequency 
channels. We confirm the expectation that {\it Planck} will detect around 
$5\times 10^4$ clusters, though the exact number is sensitive to the choice of 
several 
parameters including the baryon fraction, and also to the cluster density 
profile, so that either cosmology may predict more clusters. We also find that
the majority of detected sources should be at $z<1.5$, and we estimate that  
around one per cent of clusters will
be spatially resolved by Planck, though this has a
large uncertainty.
\end{abstract}


\section{Introduction}
\label{sec:intro}

One of the most exciting techniques for surveying for galaxy clusters at 
high redshift, which are a powerful probe of cosmology, is via the 
Sunyaev--Zel'dovich (SZ) effect (Sunyaev \& Zel'dovich 1972, 1980; Rephaeli 
1995; Birkinshaw 1999). The effect is readily measured, and indeed mapped, in 
known clusters (e.g.~Carlstrom et al.~2000) and several experiments are in 
development which will be able to 
make surveys. The two main survey types are ground-based experiments probing 
small fractions of the sky at high angular resolution, including interferometers 
such as BIMA and AmiBA and single-dish experiments such as Viper and BOLOCAM, 
and 
satellite projects making all-sky surveys at lower resolutions, 
such as the {\it Planck Surveyor} satellite 
(hereafter referred to as {\it Planck}).

The SZ effect is the distortion in the spectrum of the cosmic microwave
background (CMB) as a result of the change in energy of the CMB photons
when they scatter from hot gas, usually that in galaxy clusters. The effect
is typically separated into two components: the {\it kinetic} SZ effect is the 
Doppler shift from the bulk motion of the cluster, while the {\it thermal} SZ
effect is the gain in energy that the photons experience because the gas 
is at a much higher temperature than the CMB. As well as being of interest in 
its own right, a detailed understanding of the SZ effect is necessary in order 
to remove its effect on primary cosmic microwave background anisotropies.

To date, two main approaches have been taken to model the SZ effect: 
semi-analytical and direct numerical simulation.  The former method 
ordinarily assumes that the mass distribution of haloes at a given redshift 
is accurately represented using the Press--Schechter formalism 
(Press \& Schechter 1974). Making simplifying assumptions such as 
isothermality allows the abundance of haloes as a function of the thermal 
SZ signal to be constructed (\eg Barbosa et al. 1996; Eke, Cole \& Frenk 1996).
Although Press--Schechter
is in good agreement with results from dark matter simulations over a wide range
of mass scales, it fails for the rare, most massive, objects in a way
which has now been carefully quantified (Jenkins et al.~2000), and indeed these
corrections have been incorporated in a recent study by Bartelmann (2000).
Such methods have the drawback of making simple assumptions about the structure 
of the gas within the haloes, and also cannot be used to study the 
kinetic effect since they do not allow a direct prediction of the halo peculiar 
velocity field. Although it has recently become possible to study the kinetic 
effect 
through more detailed semi-analytic techniques (e.g.~Benson et al.~2000; 
Valageas, Balbi \& Silk 2000), such treatments are extremely complex.

Numerical simulations, though much more computationally intensive, are
able to directly model the formation of structure in both baryons and
dark matter. Hydrodynamical $N$-body simulations are the natural tool; early 
studies were made by Thomas \& Carlberg (1989) and by Scaramella, Cen \& 
Ostriker (1993), and more recently the much-improved numerical technology has 
been exploited by several groups (da Silva et al.~2000a,b; 
Refregier et al.~2000; Seljak, Burwell \& Pen 2001; 
Springel, White \& Hernquist 2000; Gnedin \& Jaffe  2000). 
Such work is now reaching a state of maturity where the importance of 
additional physics is starting to be considered. These simulations have proven 
capable of making maps of the SZ effect at high resolution (around one arcminute 
in most of the quoted papers, but at arcsecond resolution in the small-scale 
simulations of Gnedin \& Jaffe 2000), but it has not been possible to make maps 
covering a large area of sky and hence capturing the rarest, brightest objects.

In this paper we take an approach which can be considered a mid-point between
Press--Schechter and full hydrodynamical simulations, which is to use the
largest existing $N$-body simulations.  These are the Hubble Volume 
simulations, run by the Virgo Consortium (Evrard et al. 2001, in preparation;
Jenkins et al. 2000; Colberg et al. 2000), for two popular cosmological 
models --- a critical-density 
model with a modified power spectrum (which we refer to as $\tau$CDM) and a 
flat, 
low-density model with a non-zero cosmological constant (which we refer to as 
$\Lambda$CDM).  The main advantage of these simulations over hydrodynamical 
simulations is their volume; typical hydrodynamical simulations are performed
using a box-length of $\sim 100 \hMpc$; the Hubble Volume simulations are over 
an
order of magnitude larger; both models contain
$N=10^{9}$ particles, within a $2h^{-1}$ Gpc (comoving) box for the $\tau$CDM 
simulation and a $3h^{-1}$ Gpc
box for the $\Lambda$CDM simulation. In particular, this allowed ``lightcones'' 
to be constructed as the simulations were performed, obviating the need to stack 
simulation boxes as in the hydrodynamical simulations approach. Hence,
the provision of lightcones allows us to measure the abundance of clusters over 
the range of appropriate redshifts, for a range of mass scales that is larger
than in any of the previous work.  Working with the simulation data directly 
avoids the need for approximate mass functions such as Press--Schechter, and 
also
gives us direct access to the kinetic effect.  However, the fact
that the simulations are dark matter only means that we have to make some of the
simplifying assumptions concerning cluster properties characteristic of
semi-analytic methods. Therefore, such large-scale simulations are appropriate 
for making predictions for all-sky surveys rather than high-resolution 
measurements, 
and so our main focus of this paper is directed towards predictions for the 
{\it Planck} satellite. As well as source counts, we will be interested in the 
expected size and redshift distributions of the clusters.

\section{Method}
\label{sec:method}

\subsection{SZ definitions}

The SZ effect produces a fluctuation in the surface brightness of the
CMB from the inverse Compton scattering of the CMB photons
off electrons within an ionized plasma. This is due to the internal
(thermal) motions of the electrons and also the bulk (kinetic) motion 
of the plasma, relative to the CMB reference frame. The total fluctuation 
can be expressed as (\eg Rephaeli 1995)
\begin{equation}
\delta I_{\nu} = I_0  \left[ g(x) y - h(x)\beta \tau \right] \,,
\end{equation}
where $I_0=2k_{{\rm B}}^3T_0^3/h^2c^2$, $T_0=2.725 \, {\rm K}$ is
the mean temperature of the microwave background (Mather et al.~1999) and
$\beta=v_r/c$ is the radial component (\ie projected along the line of sight)
of the plasma's bulk velocity in units of the speed of light in a vacuum. 
The functions $g(x)$ and $h(x)$, known as spectral functions, contain 
frequency-dependent information for the thermal and kinetic effect 
respectively, with $x=h\nu/k_{{\rm B}}T_0$. Specifically,
\begin{equation}
g(x) = {x^4 e^{x} \over (e^{x}-1)^2} \,
\left[ x {e^{x}+1 \over e^{x}-1} - 4 \right],
\label{eqn:gofx}
\end{equation}
and
\begin{equation}
h(x) = {x^4 e^{x} \over (e^{x}-1)^2}.
\label{eqn:hofx}
\end{equation} 
The dependence of $\delta I_{\nu}$ on the thermodynamic state of the gas 
is given by the Comptonization parameter, $y$, and the optical depth due to 
electron scattering, $\tau$. For the thermal effect
\begin{equation}
y = {k_{{\rm B}} \sigma_T \over m_e c^2} \, \int \, n_e T_e \, dl \,,
\label{eqn:thermalSZ}
\end{equation}
i.e.~an integral of the electron pressure along the line of sight, while for
the kinetic SZ effect we have
\begin{equation}
\tau = \sigma_T \, \int \, n_e \, dl \,.
\label{eqn:kineticSZ}
\end{equation}

The total SZ flux\footnote{We use the term ``flux'' to mean the change in flux
caused by the SZ effect relative to the mean flux of the CMB.}
 of a source is then the integral of $\delta I_{\nu}$ 
over its subtended solid angle
\begin{equation}
S_{\nu}(x) = S_0 \int \, \left[ g(x)y-h(x)\beta \tau \right] \,
d\Omega,
\label{eqn:totalflux}
\end{equation}
where $S_0 = 2.29\times 10^{4} \mJy \, {\rm arcmin^{-2}}$
($1 \mJy= 10^{-26} {\rm erg \, s^{-1} m^{-2} Hz^{-1}}$), 
assuming that the dimensions of $d\Omega$ are $\arc2$ (the bracketed terms are
dimensionless). Note that $y$ and $\tau$ are positive quantities, whereas 
$\beta$ can be positive (if the source
is moving towards the observer) or negative (if the source is moving away
from the observer). The function $g(x)$ is zero for $x \simeq 3.83$ ($\nu 
\simeq$ 217 GHz),
negative if $x<3.83$ causing a {\it decrement} in the CMB surface brightness,
and positive if $x>3.83$, causing an {\it increment}. The function $h(x)$ is 
always positive. Hence, the signal due to the SZ effect can be deconvolved
from the total CMB signal, into thermal and kinetic components if the 
signal is observed over a range of frequencies.

For the thermal effect, it is convenient to quote the frequency-independent 
part of equation~(\ref{eqn:totalflux})
\begin{equation}
Y = d_A^{-2} \, \int \, y \, dA,
\label{eqn:thermalY}
\end{equation}
where $d\Omega=dA/d_A^2$, $d_A$ is the (cosmology-dependent) 
angular diameter distance to the source and the integral is performed 
over the source's projected area. Since $y$ is dimensionless, the dimension
of $Y$ is that of a solid angle.
Hence, $Y$ represents a weighted measure of the 
solid angle of the CMB that is obscured by the given source. Note that although
$y$ is independent of cosmology (except through the expected distribution of
sources), $Y$ depends on cosmological parameters via the relation between
the (inferred) physical and (observed) angular size of the source, through 
$d_A$.

\subsection{Simulation details and object selection}

\begin{table}
\caption{Key parameters of the Hubble Volume simulations.}
\begin{center}
\begin{tabular}{ccccccc}
Model & $\Omega_0$ & $\Omega_\Lambda$ & $h$ & $\Gamma$ & $\sigma_8$ &
$m/10^{12}\hMsol$\\ 
\hline
$\tau$CDM & 1.0 & 0.0 & 0.5 & 0.21 & 0.6 & 2.22\\
$\Lambda$CDM & 0.3 & 0.7 & 0.7 & 0.21 & 0.9 & 2.25\\
\end{tabular}
\label{tab:param}
\end{center}
\end{table}

We consider both cosmological models adopted for the Hubble Volume simulations, 
labelled $\tau$CDM and $\Lambda$CDM respectively. Key parameters of the
simulations are listed in Table~\ref{tab:param}, namely the density
parameter $\Omega_0$, the cosmological constant $\Omega_\Lambda \equiv
\Lambda/3H_0^2$, the Hubble constant $h \equiv H_0/100 \kmsMpc$, the power
spectrum shape parameter $\Gamma$, the variance $\sigma_8$ of linear
fluctuations smoothed with a top-hat filter on the (comoving) scale of 
8 $\hMpc$, extrapolated to $z=0$, and the particle mass $m$.

We use the publicly-available dark matter halo catalogues from the lightcone
datasets.\footnote{These are available at {\tt www.mpa-garching.mpg.de/Virgo/}} 
Haloes are defined as overdense {\it spheres} of particles with a mean 
internal density, $\left<\rho \right>=\Delta_{c} \, \rho_{\rm cr}(z)$, 
where $\rho_{\rm cr}(z) \equiv 3H(z)^{2}/8\pi G$ is the critical density 
and $H(z)$ is the Hubble parameter. For these catalogues $\Delta_{c}=200$,
which for $\Omega=1$ is close to the canonical value of 178 predicted 
by the spherical top-hat collapse model (Lacey \& Cole 1993). However, for 
the $\Lambda$CDM model our adopted value of 200 is around a factor of 2 
larger than predicted by the top-hat model at $z=0$. For the main results 
presented in this paper, the choice of density threshold 
is only used to normalize the mass distribution within haloes and not to
explicitly define their sizes. 

The information extracted from the simulations is the mass,
redshift and velocity distributions of the dark matter haloes. The range of mass 
scales for the haloes resolved by the Hubble Volume simulations is 
nearly two orders of magnitude ($\sim 3 \times 10^{13}$ to $10^{15} \hMsol$),
covering the scales relevant to galaxy clusters. We assume that the brightest SZ 
sources
are due to the intracluster medium (ICM) trapped within the potential wells 
of cluster haloes, known to be a robust approximation from the results of full
hydrodynamical simulations.

We use the lightcone datasets specific to each cosmology to construct
two mock cluster surveys. Firstly, we utilize the MS sphere dataset, 
covering the full $4 \pi$ steradians from $z=0$ to $z=0.44$ for 
$\tau$CDM and $z=0.58$ for $\Lambda$CDM. Beyond these redshift
limits, we graft on both PO \& NO octant datasets, spanning 
a total of $\pi$ steradians between $z=0.44$ ($z=0.58$) and
$z=1.25$ ($z=1.44$) for $\tau$CDM ($\Lambda$CDM).\footnote{We note that
there is a small degree of overlap between the PO \& NO datasets at
the highest redshifts. However, removing one of these datasets does
not significantly affect our results and so we include both for the sake of 
completeness.} Finally, we use the DW datasets which cover 100 square degrees 
and 
extend to $z=4.37$ for $\tau$CDM and
$z=4.6$ for $\Lambda$CDM. Combining the datasets allows us to have
a mass-limited survey beyond the redshifts where the first resolved
structures in the simulations form. 

We summarize key properties of each dataset used in 
Table~\ref{tab:lightcone}. Columns~1 \& 2 give the cosmological model
and dataset label respectively; Columns~3 \& 4 list the redshift
limits used in the surveys; Column 5 gives the solid angle of each
dataset and Column 6 lists the number of objects extracted from
each dataset.

\begin{table}
\caption{Key properties of the lightcone datasets.}
\begin{center}
\begin{tabular}{ccccll}
Model & Dataset & $z_{\rm min}$ & $z_{\rm max}$ & $\Omega$/sr & $N_{\rm obj}$\\
\hline
$\tau$CDM & MS & 0.00 & 0.44 & $4 \pi$ & 1136333\\  
$\tau$CDM & NO & 0.44 & 1.25 & $\pi/2$  & 227525\\
$\tau$CDM & PO & 0.44 & 1.25 & $\pi/2$  & 225820\\
$\tau$CDM & DW & 1.25 & 4.37 & $0.03$  & 199\\
\\
$\Lambda$CDM & MS & 0.00 & 0.58 & $4 \pi$ & 1539281\\  
$\Lambda$CDM & NO & 0.58 & 1.46 & $\pi/2$  & 625138\\
$\Lambda$CDM & PO & 0.58 & 1.46 & $\pi/2$  & 612549\\
$\Lambda$CDM & DW & 1.46 & 4.60 & $0.03$  & 2689\\
\end{tabular}
\label{tab:lightcone}
\end{center}
\end{table}

\subsection{Baryon distribution}

Given the mass and redshift of each halo, calculating
the resulting SZ fluxes requires knowledge about the thermodynamic 
(both density and temperature) state of intracluster gas.
Although the dimensions of the simulations are ideal for our purposes,
they did not explicitly model a baryonic component. Therefore, we need 
to make some of the same assumptions made in analytical calculations.
The first of these is to define the relationship between the electrons
and baryons, and for this, we assume that the ICM is fully ionized,
so that $n_e = 0.88 \, \rho/m_{\rm H}$ (assuming a helium mass fraction,
$Y=0.24$), where $m_{\rm H}$ is the mass of a hydrogen atom and $\rho$
is the baryon mass density.

To relate the baryonic mass to the total mass of each cluster, we assume
a global baryon fraction, $f_{\rm b}=M_{\rm b}/M_{\rm tot}$. We neglect
the presence of cold gas and stars, which only reduces the gas fraction by
around 20 per cent for the mass range of clusters relevant to this study 
(e.g. Balogh et al. 2001). 
The SZ signal simply scales proportional to $f_{\rm b}$, so theoretical 
predictions can readily be rescaled to a different value (though 
ultimately there are nontrivial effects in deciding whether extended 
sources are detectable against a background). As emphasized recently by 
Bartelmann (2000), the choice of baryon fraction can make a significant 
difference when contrasting cosmologies. The two main observational 
constraints of $f_{\rm b}$ are from X-ray observations of clusters
(e.g.~White et al.~1993; Ettori \& Fabian 1999) and from primordial 
nucleosynthesis calculations (e.g. Burles \& Tytler 1998), assuming
$f_{\rm b}=\Omega_{\rm b}/\Omega_0$ and that the mix in clusters
represents the cosmic mean. Both determinations are concordant with a
low-density cosmology but discrepant if there is a critical density 
(the so-called baryon catastrophe). Therefore, it 
appears that if critical density is to have any chance of fitting the 
whole range of available observations, then the baryon density must be 
higher than nucleosynthesis allows, and we note in passing that there 
are some moderately well motivated modifications of standard 
nucleosynthesis which make the element 
abundances compatible with a higher baryon density (see e.g.~Kang \& Steigman 
1994; Lesgourgues \& Peloso 2000; Kaplinghat \& Turner 
2001). We therefore adopt the observed baryon fraction from clusters when 
plotting results, taking the value from Ettori \& Fabian (1999) of $f_{\rm 
b}=0.06 \, h^{-3/2}$. This gives $f_b = 0.10$ for the low-density case and 
$f_b = 0.17$ for the high-density case.
The former is in reasonable agreement with the preferred 
standard nucleosynthesis value $\Omega_{\rm b}=0.019h^{-2}$ (Burles \& Tytler 
1998). In the critical-density case our main results will assume the cluster 
baryon fraction number, but occasionally we will show results based instead on 
the nucleosynthesis value $f_b = 0.076$ for comparison.

\subsection{Cluster temperatures}

To evaluate the integral in equation~(\ref{eqn:thermalSZ}) requires knowledge
of the electron pressure profile of the cluster. To simplify the discussion,
we assume that the ICM can be modelled as isothermal for all but the outer
regions of clusters. $T_e$ can then be removed from the integral in 
equation~(\ref{eqn:thermalSZ}), and in equation~(\ref{eqn:totalflux}) the 
thermal component scales proportional to the total energy of the cluster
and the kinetic as the cluster mass. Hydrodynamical simulations have confirmed
that radial variations in the ICM temperature are weak, being within a 
factor of two out to the virial radius 
(e.g. Eke, Navarro \& Frenk 1998; Pearce et al. 2000). 
We have calibrated temperature to mass using the relation
\begin{equation}
M_{200} = 2.5\times 10^{13} \, (T_{\rm x}/{\rm keV})^{1.5} \, (1+z)^{-3/2} \, 
\hMsol \,,
\label{eqn:mtrel}
\end{equation}
where $M_{200}$ is the mass within each halo where the mean internal density,
$\left<\rho\right>=200\rho_{\rm cr}$, as defined above. 
Equation~(\ref{eqn:mtrel})
is consistent with the observed relation derived by Horner, Mushotzky \& 
Scharf (1999), using spatially-resolved {\it ASCA} X-ray temperature data
for a cluster sample with $z<0.1$. 
(Note that we assume equivalence between the 
observed emission-weighted temperatures and the intrinsic electron temperature
of the gas.) To account for non-zero redshift scaling,
we assume that $T \propto M_{200}/R_{200}$ (the virial relation) and 
\begin{equation}
R_{200} = \left({2 G \over 200} \, {M_{200} \over H^{2}(z)}\right)^{1/3}. 
\label{eqn:R200}
\end{equation}
We note that  $H^{-2/3}(z) \propto \Omega^{1/3}(z)/(1+z)$ and neglect the
weak dependence on $\Omega$. This results in the presence of the
$(1+z)^{-3/2}$ term on the right-hand side of equation~(\ref{eqn:mtrel}).

\subsection{Density profiles}
\label{subsec:sizes}

For our main results, we consider two forms for the distribution of baryons 
within cluster potential wells, since the precise form of $\rho(r)$ is not yet 
well determined from X-ray observations (although significant improvements 
should come from {\it Chandra} and {\it XMM--Newton} observations). 
Historically,
cluster ICM density profiles are obtained by fitting a $\beta$ model to
the (azimuthally-averaged) observed surface brightness distribution 
and making an assumption regarding the temperature distribution (usually
isothermality). However, most of these observations suffer from the inability 
to probe the outer radii of clusters because of the high X-ray background 
(which is a combination of instrumental noise and the intrinsic X-ray background 
of 
unresolved sources), particularly in low temperature systems. 
Consequently, the outer slope of the density profile is usually ill-constrained.
In light of these difficulties, we choose to consider two 
density profiles for the gas that are already well documented in the literature.
The first of these is the $\beta$ model profile, the standard model
for fitting the observed ICM surface brightness distributions outwith cooling 
flow regions
\begin{equation}
\rho_{\beta}(r) = 	{\rho_{\beta}(0) \over
			[1+(r/r_{{\rm c}})^{2}]^{-3\beta/2}},
\label{eqn:betaprof}
\end{equation}
where the parameters $r_{{\rm c}}$ and $\beta$ determine the core radius and 
the asymptotic slope of log($\rho$) at large radii respectively. We
calculate $\rho_{\beta}(0)$ assuming that the integrated baryonic mass 
of the cluster should equal $f_{\rm b} \, M_{200}$ at $R_{200}$. 
For the core radius, we set $r_{{\rm c}}=0.07 \, R_{200}$  
(e.g. Lloyd--Davies, Ponman \& Cannon 2000). 
Measured $\beta$ values for individual clusters are typically between 0.5 and 
0.8 
(e.g. Jones \& Forman 1984,1999; Horner et al. 1999; 
Mohr, Mathiesen \& Evrard 1999; see also Vikhlinin, Forman \& Jones 1999). 
For the results presented in this paper we adopt $\beta=5/6$ ($\sim 0.83$) 
inferring an asymptotic slope at large radii of $-2.5$. Although this is rather
high, we note that our assumption of isothermality is problematic when 
calculating $y$ for shallow density profiles, since the temperature should drop 
significantly outside the virial radius (this is verified by simulations). 
Therefore, we choose to interpret our choice of $\beta$ as a way in which to 
parameterize the radial fall-off of the ICM pressure, $\rho \, T$, at large
radii. 

The second density profile considered in this paper is well-motivated by results 
from numerical simulations: the NFW profile (Navarro, Frenk \& White 1997), 
which 
provides a good 
fit to average dark matter halo density profiles over a wide range of radii, for 
all popular 
CDM and scale-free cosmological models
\begin{equation}
\rho_{\rm NFW}(r) =	{\rho_{\rm NFW}(0) \over
			(r/r_s)(1+r/r_s)^2}.
\label{eqn:nfwprof}
\end{equation}
The NFW model contains just one adjustable parameter, known as the concentration
parameter, $c=R_{200}/r_s$ and determines the radius at which the density 
profile steepens from a slope of $-1$ to an outer slope of $-3$. We normalize 
the 
profile as for the $\beta$ model and calculate $c$ as a function of halo mass 
and 
redshift using the method described in Appendix A of Navarro et al.~(1997). 

\vspace{0.5cm}
To summarize, the inner regions of clusters are modelled as isothermal with
either a $\beta$ model (constant core) or NFW (cuspy) density profile. In the
outer regions of clusters, the NFW profile is sufficiently steep that variations
in temperature are unimportant, whereas a value of $\beta$ has been chosen in
order to represent what we anticipate as the {\it weakest} dependence of $y$
with radius. Hence, we argue that the two models should span a reasonable 
distribution of what the true SZ surface brightness profiles are likely to be.

\subsection{SZ Fluxes and object sizes}
\label{subsec:fluxsize}

We concentrate on calculating fluxes for 3 frequencies:
143, 217 and 353 GHz. These frequencies correspond to the central
values of 3 channels on the {\it Planck} HFI detector (Puget et 
al.~1998). Although there are other channels at higher frequencies
(545 and 857 GHz), the signal at those frequencies will be dominated by dust 
emission.
These frequencies have been specifically chosen in order to allow the best 
separation of the
signal due to the SZ effect and the true primordial  
temperature fluctuations. 
143 GHz corresponds to the maximum {\it decrement} in the CMB spectrum
that is induced by the thermal SZ effect. At 217 GHz, there is no signal 
due to the thermal SZ effect, allowing the contribution from the kinetic 
effect to be measured. 353 GHz corresponds to the maximum {\it increment}
in the CMB spectrum induced by the thermal SZ effect.

The {\it Planck} HFI detector will not measure temperature fluctuations only
at these precise frequencies. Rather, each channel will have its own
specific frequency response function. We model each response function as a 
Gaussian and assume that the full-width half-maximum (FWHM), 
$2(\nu-\nu_0) / \nu_0=0.37$ (Hobson et al.~1998), where $\nu_0$ is the 
central frequency. We calculate the average fluxes that are measured for
each central frequency by replacing $g(x)$ (equation~\ref{eqn:gofx}) and
$h(x)$ (equation~\ref{eqn:hofx}) with 
\begin{equation}
G(x_0) = {1 \over \sqrt{\pi}\sigma} \, \int_{0}^{\infty} \, g(x) \, 
\exp \left(-{(x-x_0)^2 \over \sigma^2} \right) \, dx,
\label{eqn:biggofx}
\end{equation}
and
\begin{equation}
H(x_0) = {1 \over \sqrt{\pi}\sigma} \, \int_{0}^{\infty} \, h(x) \, 
\exp \left(-{(x-x_0)^2 \over \sigma^2} \right) \, dx,
\label{eqn:bighofx}
\end{equation}
where $\sigma=0.37x_0/2\sqrt{\ln 2}$.
For an ideal detector of width $\nu-\nu_0 \rightarrow 0$,
(i.e. measuring the flux at a single frequency), the response function 
would approach the Dirac delta function and $g(x)$ and $h(x)$ would both
be recovered.

The simplest model used to construct SZ fluxes is to assume 
that the size of each halo is $R_{200}$, in which case the total flux 
(equation~\ref{eqn:totalflux}) does not depend on the thermal
distribution of the intracluster gas. For the thermal SZ
effect, we can define the integrated $y$ out to $R_{200}$, $Y_{200}$, 
by combining equations~(\ref{eqn:thermalSZ}), (\ref{eqn:thermalY})
and (\ref{eqn:mtrel}) to obtain
\begin{equation}
Y_{200} = 1.7 \times 10^{-3} {f_{\rm b} \over h} 
\left({M_{200}\over 10^{14}}\right)^{5/3}
\left({d_A \over 500}\right)^{-2}
(1+z) \, {\rm arcmin}^2 \,,
\label{eqn:Y200}
\end{equation}
where $M_{200}$ is in units of $\hMsol$ and $d_A$ is in units of $\hMpc$. 
This model assumes that there is negligible signal from haloes outside
$R_{200}$, which would be the case if the gas is only ionized inside $R_{200}$
(i.e. with a temperature significantly greater than $10^{4}$K). 
However, this approach does not allow us to account for beam 
convolution, which requires the exact distribution of the gas to be modelled,
and so we choose to present this model only to contrast with our main 
results.

The total SZ flux from each cluster is calculated
by first defining a beam-convolved $y$ profile, assuming that the 
{\it Planck} beam profile is adequately represented by a Gaussian
\begin{equation}
y_c(\theta) = \int y ( \mbox{\boldmath{$\theta$}}^{'} ) \, 
b(|\, \mbox{\boldmath{$\theta$}}-\mbox{\boldmath{$\theta$}}^{'}|) \, 
d^2\mbox{\boldmath{$\theta$}}^{'} \,,
\label{eqn:yconvolve}
\end{equation}
where $\theta=|$\mbox{\boldmath{$\theta$}}$|$ is the projected angular 
radius of the cluster and
\begin{equation}
b(|\mbox{\boldmath{$\theta$}}-\mbox{\boldmath{$\theta$}}^{'}|) = 
{1 \over \pi \sigma^2} \, \exp \left(- {|\, \mbox{\boldmath{$\theta$}}^{'}-
\mbox{\boldmath{$\theta$}}^{'}|^2\over \sigma^2}\right). 
\label{eqn:beamfunction}
\end{equation}                                                     
Standard deviations are calculated using the full-width 
half-maxima quoted in the literature, \ie approximately 8 arcminutes
for the 143 GHz channel and 5 arcminutes for the 217 and 353 GHz channels.
The total SZ flux is then  
\begin{equation}
S_c(x_0) = S_0 \int_0^{\theta_{\rm bg}}
2\pi \theta  \left[ G(x_0)y_c(\theta)-H(x_0)\beta \tau_c(\theta)
\right] \, d\theta,
\label{eqn:flux_convolve}
\end{equation}
where 
the beam-convolved optical depth, $\tau_c \propto y_c/T$ in the isothermal 
limit.
The upper limit in equation~(\ref{eqn:flux_convolve}), $\theta_{\rm bg}$ is 
defined as the angular radius where $y_c$ equals the background due to 
unresolved sources, $y_{\rm bg}$. Values of $y_{\rm bg}$ are not easily 
established and one must account for not only the mean $y$, $\left<y\right>$ 
but also fluctuations about $\left<y\right>$ , which will depend on the 
beam-size. At present, only an observed upper limit exists to $\left<y\right>$,
of $\left<y\right> < 1.5\times 10^{-5}$ from the {\it FIRAS} instrument on the 
{\it COBE} satellite (Fixsen et al.~1996), while fluctuations 
about $\left<y\right>$ are constrained only on very large angular scales. 
However the background can be predicted from both the Press--Schechter 
approach (Bartelmann 2000) and from hydrodynamical simulations 
(da Silva et al.~2000b). We adopt values from the latter and choose 
to define the background, $y_{\rm bg}$  as the 3 sigma fluctuation above 
$\left<y\right>$. This value is approximately $y_{\rm bg}=5 \times 10^{-6}$ 
for a Gaussian beam with a FWHM of 5 arcminutes.

Finally, to estimate the sizes of objects that will be observed by
{\it Planck}, we cannot simply take the $\theta_{\rm bg}$ values, since
this does not give information regarding whether a source is either
a point source (which may be extended due to the beam convolution) or
actually resolved resolved by the beam. Instead, we define the angular
size of each object, $\theta_s$, to be the angular radius at which 
$y_c(\theta) = y_c(0)/2$, i.e. the half-width half-maximum. For objects
where $\theta_s>\theta_{\rm bg}$ (i.e. $y_{\rm bg}>y_c(0)/2$), 
then $\theta_s$ is set to $\theta_{\rm bg}$.
We define sources with $\theta_s$ greater than the beam FWHM as resolved,
since this will produce flux in more than one detector pixel (which
will typically have a size of order the beam FWHM).
\section{Results}
\label{sec:results}

\subsection{Halo distributions}
\label{subsec:halodist}

\begin{figure}
\centering
\centerline{\psfig{file=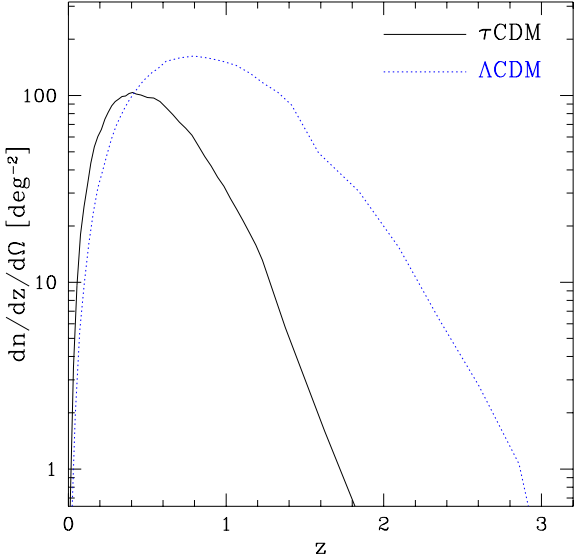,height=8cm}}
\caption{Redshift distributions of haloes from each survey.}
\label{fig:nz}
\end{figure}

Fig.~\ref{fig:nz} shows the number of haloes per unit redshift per square
degree, as a function of redshift, for all haloes extracted from both
simulations.  Firstly, the area under the $\Lambda$CDM redshift distribution is
greater than for $\tau$CDM due to the larger number of objects in the
$\Lambda$CDM catalogue ($\sim 2.8\times 10^6$, as opposed to $\sim 1.6\times
10^6$).  Secondly, the peak of the $\Lambda$CDM distribution is higher than for
the $\tau$CDM case.  For example, the $\tau$CDM simulation predicts that the
number of objects per unit redshift is around 100 per square degree, whereas the
$\Lambda$CDM simulation predicts that the peak number is around 50 per cent
larger.  Secondly, the peak of the $\tau$CDM distribution occurs at $z \sim
0.4$, whereas the peak of the $\Lambda$CDM distribution occurs at $z \sim 0.8$,
resulting in a much broader redshift distribution.  Consequently, the $\tau$CDM
survey extends to $z \sim 2$ whereas the $\Lambda$CDM survey extends to $z \sim
3$.  These differences arise due to the slower rate of structure formation in
the low-density case, which requires a larger normalization of density
fluctuations (through $\sigma_8$) in order to produce the same $z=0$ abundance
of rich clusters as in the critical-density model.

\begin{figure}
\centering
\centerline{\psfig{file=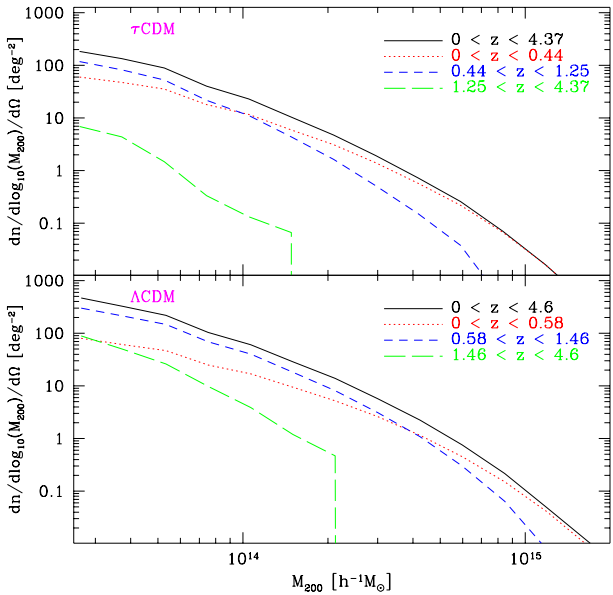,height=8.5cm}}
\caption{Halo mass functions for the $\tau$CDM survey (top panel)
and $\Lambda$CDM survey (bottom panel). The solid line is for the
combined datasets and the dotted/dashed lines illustrate contributions
from individual datasets covering various redshift intervals.}
\label{fig:dmf}
\end{figure}         

Differential mass functions of haloes are illustrated in 
Fig.~\ref{fig:dmf} for $\tau$CDM (top panel) and for $\Lambda$CDM 
(bottom panel). The solid line in each figure illustrates the 
mass function for the complete survey, in addition to contributions from
the MS dataset (dotted line), the combined PO \& NO dataset (short-dashed line)
and the DW dataset (long-dashed line). Hence, each component illustrates the
contribution from 3 redshift intervals.  The majority of high-mass objects 
($> 10^{15} \hMsol$) are at low redshift and the majority of haloes
at lower masses come from intermediate redshift. This is due to the 
characteristic halo mass (known as $M^*$, whose value roughly coincides
with the {\it knee} of the halo mass function) decreasing with increasing 
redshift, as is the case in all hierarchical models. 

We also note that the haloes with redshifts above $z \sim 1.2-1.5$ 
(\ie from the DW dataset) make only a small contribution to the total 
mass function in either model. The amplitude of the DW mass function is at 
least a factor of 5 lower than the total mass function in the 
$\Lambda$CDM case, and over a factor of 20 lower in the $\tau$CDM case, 
where structure evolves more rapidly with redshift. 
                                                   
\begin{figure}
\centering
\centerline{\psfig{file=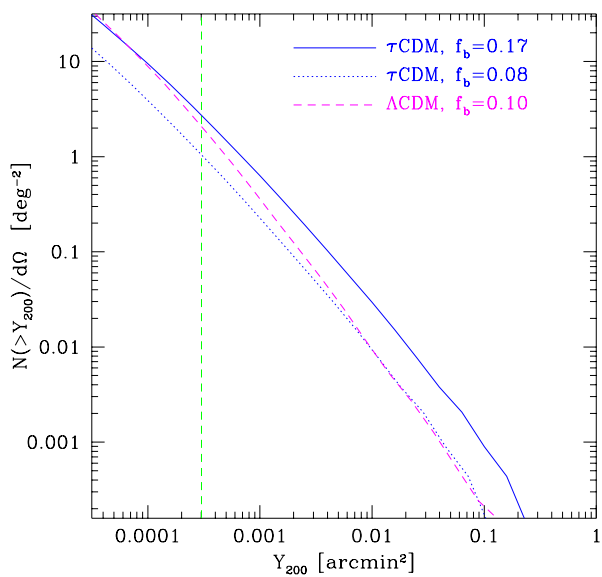,height=8cm}}
\caption{The number of haloes above a given $Y_{200}$ (defined in
the text) for the two surveys. As well as the values obtained assuming the 
cluster baryon fraction, we show the result for the $\tau$CDM 
survey with a baryon fraction around a factor of two smaller,
consistent with standard nucleosynthesis. The vertical dashed line corresponds
to the sensitivity limit of {\it Planck} ($Y_{\rm lim}\sim 3\times 10^{-4} 
\arc2$).}
\label{fig:ny}
\end{figure}

The cumulative number of haloes above $Y_{200}$ is plotted in 
Fig.~\ref{fig:ny}.
We note that our datasets are mass-limited and not flux-limited, as is the 
case with real surveys, and so if we choose to consider the full mass range 
of haloes, 
we may be incomplete in flux due to the absence of smaller mass sources that 
are sufficiently close to $z=0$ to have $Y$ values comparable to our 
resolved sources. As noted by Bartelmann (2000), a conservative 
sensitivity limit of forthcoming {\it Planck} surveys is around 
$Y_{\rm lim}=3\times 10^{-4} \arc2$. By considering only haloes with 
$Y_{200}>Y_{\rm lim}$, we find that 90 per cent of these objects
have masses greater than 25 particles for $\tau$CDM and 40 particles for 
$\Lambda$CDM respectively (the mass limit is equivalent to 12 particles). 
Hence, our constructed surveys should be sufficient to describe the 
distribution of objects with $Y>3\times 10^{-4} \arc2$. The vertical 
dashed line in Fig.~\ref{fig:ny} denotes $Y_{\rm lim}$.

The solid and dashed lines in Fig.~\ref{fig:ny} represent the
source counts above a given $Y_{200}$ for the $\tau$CDM and 
$\Lambda$CDM surveys respectively, assuming a value of $f_{\rm b}$ 
calculated from clusters. The number of sources declines more 
rapidly for the $\Lambda$CDM case than for the $\tau$CDM case. 
For $Y_{200}>10^{-4}\arc2$ both surveys predict approximately 10 
clusters per square degree, but for $Y_{200}>0.1 \arc2$ the 
$\tau$CDM survey predicts around an order of magnitude more objects 
than the $\Lambda$CDM survey. This is because the most massive 
clusters in the $\Lambda$CDM model are, on average, at higher 
redshift than those in the $\tau$CDM model. Furthermore, the angular 
diameter distance at fixed redshift is larger 
for the $\Lambda$CDM model than for the $\tau$CDM model. These two 
differences conspire to produce smaller $Y_{200}$ values for the largest 
clusters in the $\Lambda$CDM model than in the $\tau$CDM model. However, 
for lower $Y_{200}$ values, a larger contribution comes from
high redshift, where the space density of clusters is higher in the
$\Lambda$CDM model (beyond $z \sim 0.4$ -- see Fig.~\ref{fig:nz}), 
causing the $\Lambda$CDM curve to rise more steeply than the $\tau$CDM
curve.

As noted in Bartelmann (2000), the number of sources above a given $Y$ is
sensitive to the adopted value of the baryon fraction. To illustrate this,
we have also plotted the $\tau$CDM counts with a lower baryon fraction, 
which is in conflict with cluster measurements but consistent with 
nucleosynthesis determinations. As stated before, using 
$\Omega_{\rm b}=0.019h^{-2}$ (Burles \& Tytler 1998) and taking 
$h=0.5$ implies $f_{\rm b}=0.076$. Since $Y$ scales linearly
with $f_{\rm b}$, the factor of $\sim 2$ difference between the baryon
fractions scales the counts horizontally by the same factor. However,
from Fig.~\ref{fig:ny} the cumulative number of sources scales approximately 
proportional to $Y_{200}^{-4/3}$ for the $\tau$CDM case. The change 
in the predicted counts when approximately 
doubling $f_{\rm b}$ is an increase by a factor of $\sim 2.7$. Hence, the
degeneracy between $f_{\rm b}$ and $\Omega_0$ implies that the 
projected source counts is not a good constraint on one without 
accurate prior knowledge of the other. This is especially true because the 
counts are also expected to be quite sensitive to the value of $\sigma_8$.

\begin{figure}
\centering
\centerline{\psfig{file=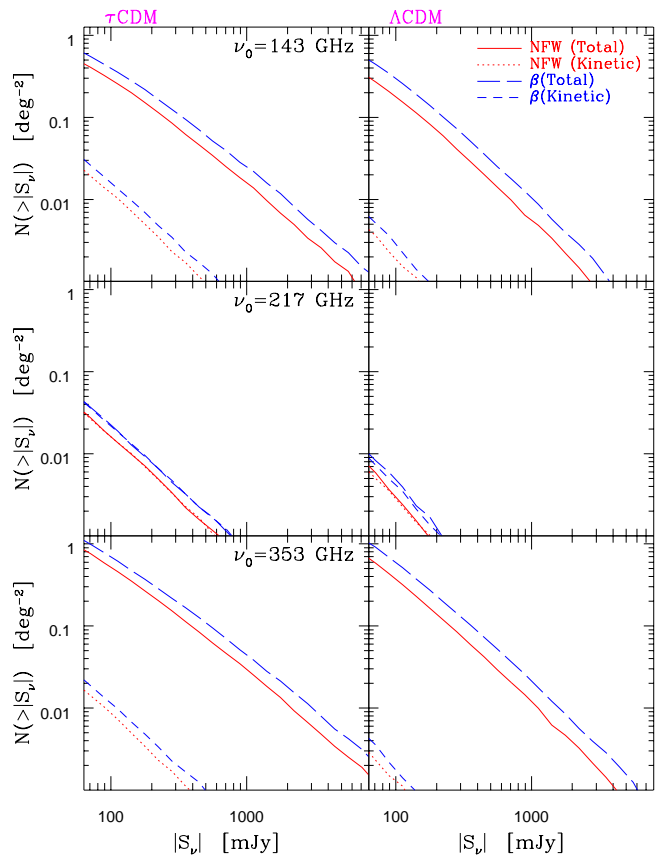,height=12cm}}
\caption{Source counts above a given absolute flux, in mJy, 
for the $\tau$CDM cosmology (left column) and $\Lambda$CDM
cosmology (right column). The rows from top to bottom specify
results for central frequencies of 143, 217 and 353 GHz respectively.
Within each panel, results are plotted for the NFW model 
(solid lines correspond to total absolute fluxes and dotted
lines correspond to the kinetic fluxes) and $\beta$ model
(long-dashed and short-dashed lines correspond to total and kinetic
fluxes).}
\label{fig:counts}
\end{figure}

\subsection{Source counts}

The number of sources per square degree above a given absolute flux,
in mJy, is plotted in Fig.~\ref{fig:counts} for two models: one 
assuming an NFW density profile and one assuming the $\beta$ model 
density profile (we will subsequently refer to these models as ``NFW'' 
and ``$\beta$''). The left column illustrates results
for the $\tau$CDM models and the right column for $\Lambda$CDM models.
From top to bottom, results are for central frequencies of 
143, 217 and 353 GHz respectively, assuming a beam FWHM of 8 arcminutes
for the 143 GHz results and 5 arcminutes for the 217 and 353 GHz results.
Within each panel, the solid and 
dotted curves are for the NFW model, illustrating counts for the total 
(thermal plus kinetic) flux and the kinetic component respectively.
Long and short dashed curves are results for total and kinetic
fluxes for the $\beta$ model.

Comparing the total flux counts between the $\tau$CDM and $\Lambda$CDM 
results for the same density profile at 143 GHz, the same trend is apparent
as in Fig.~\ref{fig:ny}: at low fluxes both models predict similar source counts 
(e.g. 
the $\beta$ model predicts $0.3-0.4$ sources per square degree above 100 mJy), 
but at higher fluxes the $\tau$CDM model predicts more sources than the 
$\Lambda$CDM model. Using the nucleosynthesis determination of
the baryon fraction for both models would lower the $\tau$CDM counts
relative to the $\Lambda$CDM counts.

At 217 GHz, the signal is completely dominated by the kinetic effect,
since at this frequency the thermal effect gives a negligible contribution
even when accounting for the finite width of the detector's response
function. The number of sources above a given flux in this channel
is at least an order of magnitude lower than in the other two channels
(which can also be seen for a given frequency, when comparing total
to kinetic flux), because the distortion caused by the kinetic effect is weaker
than the thermal effect. As a result, the brightest sources observed at 
143 and 353 GHz have their fluxes substantially reduced at 217 GHz. 
Furthermore, since the difference between the $\tau$CDM and $\Lambda$CDM 
models is greatest for the largest $Y$ values, the difference 
in source counts between the two cosmologies at 100 mJy is greater
than at the other frequencies.

For the 353 GHz channel, the source counts are higher at constant
total flux than the other two channels. Comparing this to the 143
GHz results (i.e. where the signal is also dominated by the thermal
effect), two differences cause the rise. First, the beam FWHM
has decreased from 8 arcminutes to 5 arcminutes which allows more
sources to be resolved. Second, for the constant bandwidth 
($\Delta \nu/\nu=0.37$), the dispersion of the beam $\sigma$ is 
proportional to $\nu$ and so the SZ flux is higher at 353 GHz than 
at 143 GHz.

At all frequencies and in both cosmologies, the $\beta$ model predicts
more sources above a given flux than the NFW model. Since the $\beta$ 
model density profile is shallower at large radii than the
NFW profile, corresponding fluxes in the former model are larger for 
individual objects which consequently increases the total number of 
sources above a fixed flux.

\subsection{Redshift distributions}

\begin{figure}
\centering
\centerline{\psfig{file=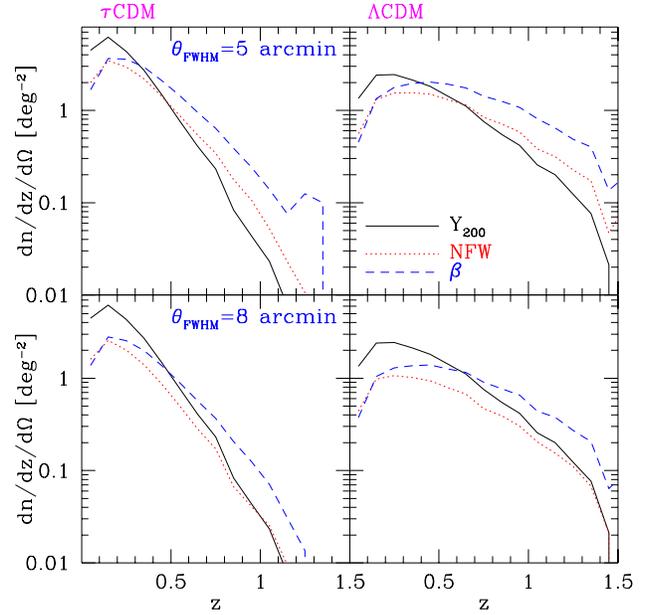,height=8.5cm}}
\caption{Redshift distributions of haloes in the $\tau$CDM
survey (left column) and $\Lambda$CDM survey (right column),
with $Y>3\times 10^{-4} \arc2$. Each panel illustrates results 
for 3 different estimates of $Y$, labelled as $Y_{200}$, NFW \& $\beta$ 
(see text for details). Results for the latter two cases were 
generated assuming a Gaussian beam function with a FWHM of 5
arcminutes (top panels) and 8 arcminutes (bottom panels).}
\label{fig:dndzY}
\end{figure}

Redshift distributions of SZ sources are given in Fig.~\ref{fig:dndzY},
which is split into 4 panels. The panels on the left of the figure 
illustrate results for the $\tau$CDM survey while the panels on the right 
illustrate results for the $\Lambda$CDM survey. Each sample only contains
haloes with $Y>Y_{\rm lim}$. The solid lines represent the redshift 
distributions of haloes assuming $Y=Y_{200}$, with no prior assumption 
of how the gas is distributed within $R_{200}$ (equation~\ref{eqn:Y200}).
Dotted and dashed lines correspond to distributions for the NFW and
$\beta$ models respectively. The top panels assume a 
beam FWHM of 5 arcminutes and the bottom panels assume a beam FWHM of 
8 arcminutes. (In practice the limit for detection of SZ clusters comes from 
combining different channels which have different FWHM, and we expect the 
effective beam profile for SZ detection by {\it Planck} to lie between these 
values.)

We first consider the $Y_{200}$ redshift distributions, which assume
no beam convolution and in effect assume all clusters are unresolved by 
{\it Planck} and contribute signal only from their inner regions. The amplitude 
of the peak in the redshift 
distributions for both cosmological models is over an order of magnitude 
lower than in Fig.~\ref{fig:nz}, where haloes of all masses are
considered. As is evident from Fig.~\ref{fig:ny}, there is a significant 
number of sources with $Y<Y_{\rm lim}$, and so the deficit in clusters
in Fig.~\ref{fig:dndzY} is due to the removal of these sources, and this 
confirms the near completeness of the sample.
The peak redshift values for the $Y_{200}$ distributions
are $z \sim 0.1$ for $\tau$CDM and $z \sim 0.4$ for 
$\Lambda$CDM respectively, which are also considerably lower than when 
the full range of haloes are considered. Since 
$Y_{200} \propto M_{200}^{5/3} (1+z) / d_A^{2}(z)$, the imposed lower 
limit in $Y_{200}$ implies a lower mass limit which increases with 
$z$, due to the effect of $d_A$. Hence, this reduces the number of clusters
by a larger factor at high $z$ than at low $z$, causing the peak of the 
distribution to move to a lower redshift. Note in particular that the height of 
the $\Lambda$CDM redshift distribution is now lower than in the 
$\tau$CDM redshift distribution, by around a factor of 2.5, since the
peak in the original $\Lambda$CDM sample is at a higher redshift than
in the $\tau$CDM sample. However, at higher redshift (beyond $z \sim 0.4$) the 
$\Lambda$CDM model
still predicts an excess of sources compared to the $\tau$CDM model.

Comparing the $Y_{200}$ redshift distributions to those produced by the NFW
model (in the top panels of Fig.~\ref{fig:dndzY}, assuming a beam FWHM
of 5 arcminutes), the NFW model  predicts fewer clusters per square
degree than the $Y_{200}$ model for redshifts below $\sim 0.5$ 
for $\tau$CDM ($\sim 0.7$ for $\Lambda$CDM) but more clusters beyond 
these redshifts. There are several reasons for the difference.

Firstly, the size of each cluster for the NFW model 
(\ie the upper limit of the $Y$ integral) is no longer fixed at $R_{200}$ 
but is determined by angular radius at which $y_c=y_{\rm bg}$ 
(i.e. $\theta_{\rm bg}$). This allows clusters with sizes 
$\theta_{\rm bg}$ that are less or greater than $\theta_{200}$ 
to be included in the sample, providing 
they have $Y_{\rm NFW}>Y_{\rm lim}$. Note also that since $y$ is an
integral along the line of sight, the NFW model does not calculate $y_c$ 
assuming that each halo has a diameter $2R_{200}$ in this direction;
rather, the limits are selected such that the integrated $y$ has converged.

Secondly, for a halo with constant $M_{200}$ the value of $y$ increases
at fixed radius with redshift, because the density profile has a higher 
normalization at higher redshift. This is due to the definition of each
halo being a region of large density contrast (i.e.~on average 
200 relative to critical). As redshift increases, the critical density 
increases as $\Omega^{-1}(z)\, (1+z)^3$, causing an increase in the
average physical density of the halo. This causes the radius at which
the integrated $y$ reaches the background to grow with redshift,
increasing the value of $Y$.

Thirdly, when the $y$ profile is convolved with the beam function, 
it is flattened out to a radius approximately equivalent to the 
angular size of the beam. As redshift is increased, the apparent
angular size of the cluster decreases and so the convolution has
a significant effect on the profile out to large physical radii.

Hence, the deficit of clusters in the NFW model at low $z$, when
compared to the $Y_{200}$ model, is due to low-mass objects which are not
resolvable against the background when convolved with the beam
($Y_{\rm NFW}<Y_{\rm lim}$). However, at higher redshifts the net
effect is that the profiles become higher relative to the background 
(even though the convolution has a greater effect), leading to a larger
angular size and consequently, for a halo of constant $M_{200}$, increasing
$Y_{\rm NFW}$ relative to $Y_{200}$. 

The same argument can be applied to the $\beta$ model, 
albeit predicting more clusters than the NFW model for approximately 
$z>0.2$. The $\beta$ model density profile decreases more weakly at
large radii than the NFW profile: $\rho_{\beta} \sim r^{-2.5}$ for our
choice of $\beta=5/6$, whereas $\rho_{\rm NFW} \sim r^{-3}$. Hence,
$Y_{\beta}>Y_{\rm NFW}$ for all but the very lowest redshifts where
convolution effects dominate.

In the bottom panels of Fig.~\ref{fig:dndzY}, results are displayed for a 
beam FWHM of 8 arcminutes; remember that the effective resolution of 
{\it Planck} for SZ will lie between 5 and 8 arcminutes. The number of clusters 
in both the NFW and 
$\beta$ models decreases when compared to the previous results
(by less than a factor of 2), although the value of the peak redshift is not 
significantly affected. The larger beam FWHM causes the convolution to
have an effect out to larger radii, which reduces the number of objects
able to be resolved against the background.

\subsection{Cluster sizes}

\begin{figure}
\centering
\centerline{\psfig{file=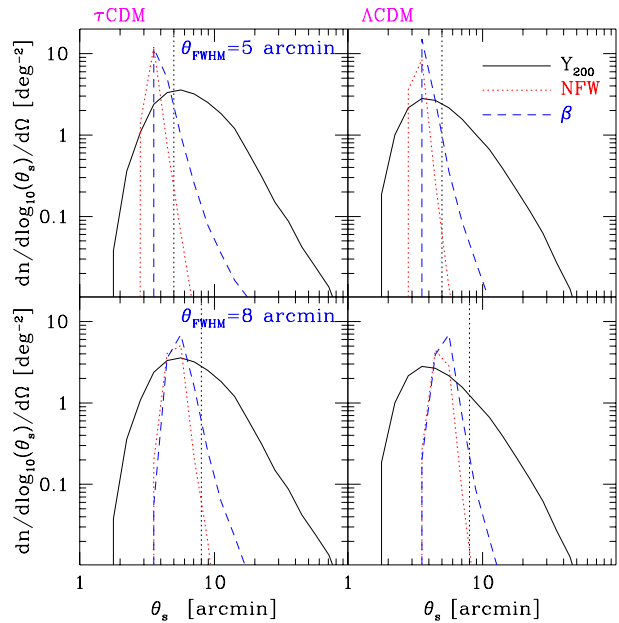,height=8.5cm}}
\caption{Angular size distributions of clusters in the $\tau$CDM
survey (left column) and $\Lambda$CDM survey (right column), with
$Y>3\times 10^{-4} \arc2$. Solid curves depict distributions assuming
$Y_{200}$ values, dotted curves are for the NFW model and dashed values
are for the $\beta$ model. The dotted vertical lines illustrate the 
value of the FWHM, which is 5 arcminutes in the top panels and 8 arcminutes
in the bottom panels.} 
\label{fig:angsize}
\end{figure}

While most clusters are effectively point sources at the {\it Planck} 
resolution, it is expected to resolve the nearer ones. In Fig.~\ref{fig:angsize} 
we plot the number of clusters, as a function of their angular size $\theta_s$. 

The $Y_{200}$ model shows the angular sizes corresponding to the clusters' 
$R_{200}$ values, $\theta_{200}=R_{200}/d_A(z)$. The number of sources peaks at 
around 5--6
arcminutes for the $\tau$CDM case and around 3--4 arcminutes
for the $\Lambda$CDM case. Both distributions are skewed such that
there are more objects with sizes greater than the peak size than 
below it. Broadly speaking, the objects with low $\theta_s$ values
are at high redshift (the minimum resolved angular size of a source 
with $Y_{200}>Y_{\rm lim}$ strongly declines with redshift until  
$z \sim 0.5$, after which it flattens off), and the objects with 
large $\theta_s$ values are at low redshift. The $\tau$CDM cosmology
predicts similar numbers of sources to the $\Lambda$CDM cosmology
at the peak of the distributions  but the former predicts more
with high $\theta_s$ values, due to the excess of sources
at low redshift relative to the $\Lambda$CDM case.

However the $R_{200}$ radii do not give a good estimate of the angular size 
detectable by {\it Planck}, for which we need to model the cluster profiles 
using $\theta_s$ as described in Section~\ref{subsec:fluxsize}. We consider beam 
convolutions of 5 and 8 arcminutes, as shown by the dotted vertical lines; 
recall that the SZ source catalogue will arise from combining the different 
frequency channels and so the effective {\it Planck} resolution in SZ is 
expected to lie between these values.
For the NFW and $\beta$ models the distributions are much narrower,
and the peak number of sources higher, than for the $Y_{200}$ model. The sharp 
cut-off at low $\theta_s$ is due
to the size at which the convolved high-redshift sources no longer
contain enough flux to be detected above the sensitivity limit. For a 5 
arcminute 
beam (top panels) this size is larger for the $\beta$ model than for the NFW
model, but for an 8 arcminute beam the two are comparable.
Thus, the smallest resolved size is sensitive to the inner shape of
the convolved profiles, determined by both the intrinsic $y$ profile 
(which is different for the NFW and $\beta$ cases) and the beam size.

The position and height of the peaks in the NFW and $\beta$ model size 
distributions indicate that the majority of objects will be unresolved
by {\it Planck}. The largest number of resolved sources is found by assuming the 
$\tau$CDM cosmology and a 5 arcminute beam; even then, the NFW model predicts 
that less than 1 per cent of the sources will be resolved, although this
rises by almost an order of magnitude for the $\beta$ model. This 
difference is caused by the more extended distribution of gas in the
$\beta$ model, increasing the number of detected sources above the background.
We conclude that without more detailed information on the cluster profiles it is 
difficult to accurately predict the fraction of sources which will be 
resolvable.

\section{Conclusions}
\label{sec:conclude}

In this paper, we have used the `lightcones' generated from the
largest dark matter simulations to date, the Hubble Volume
simulations, in order to predict the distribution of sources that will be 
observable, via the Sunyaev--Zel'dovich (SZ) effect, by the forthcoming 
{\it Planck Surveyor} mission. We have used the simulations of a 
critical-density Universe with a modified initial CDM power spectrum 
($\tau$CDM), and a flat low-density Universe ($\Lambda$CDM) to
obtain masses, redshifts and peculiar velocities of dark matter haloes
from $z=0$ to $z\sim 4$. By making simple assumptions on 
the distribution of ionized gas within these haloes, we have been able
to predict the expected distribution of SZ sources as a function of 
redshift, angular size and flux in 3 of the frequency channels relevant
to the {\it Planck} HFI detector. Our calculations involve simple
modelling of the detector response function, beam convolution 
and the presence of an unresolved background. Our main conclusions are as 
follows.

The number of SZ sources above a given $Y$ rises at a higher rate (for 
decreasing flux limit) in the $\Lambda$CDM cosmology than for the $\tau$CDM 
cosmology,
due to differing redshift distributions and the angular diameter
distance--redshift relation. The $\Lambda$CDM model produces more objects
at higher redshift than does the $\tau$CDM model, as well as the angular
diameter distance being larger for a given redshift.
However, the absolute number of sources above any given $Y$ is not itself
a good discriminator between differing cosmological models. The 
normalization is sensitive to the choice of baryon fraction 
($f_{\rm b}$), since $Y \propto f_{\rm b}$ and $N(>Y)$ falls off 
rapidly with $Y$, and a significant dependence on $\sigma_8$ is also expected.

As expected, the source counts at 217 GHz are dominated by the kinetic effect,
whereas at 143 and 353 GHz the counts are dominated by the thermal effect.  At 
217 GHz, the predicted number of sources expected to be
detected by {\it Planck} above 100 mJy is around 0.04 per square degree for the
$\tau$CDM model (with cluster baryon fraction), lowering by about a factor of 5
for the $\Lambda$CDM model.  At 143 GHz (assuming a beam FWHM of 8 arcminutes),
this rises to around 0.3 per square degree for both models, and at 353 GHz the
narrower beam size implies a value of around 1 source per square degree with
$Y>3\times 10^{-4} \arc2$ (appropriate for {\it Planck}, Bartelmann
2000).  Differences in the source counts between NFW and $\beta$ models are 
within a factor of 2.

The redshift distribution of SZ sources peaks at $z \sim 0.2$ for the
$\tau$CDM model and $z \sim 0.3-0.4$ for the $\Lambda$CDM model. At redshifts
below the peak values, the expected number of clusters is insensitive to the 
adopted gas distribution. However, for redshifts above the peak, a more extended
gas (pressure) distribution gives a larger number of sources. For our adopted
NFW and $\beta$ density profiles, the latter
implies a $y$ distribution that falls off more slowly with radius, since the 
asymptotic slope of our adopted $\beta$ density profile is $r^{-2.5}$, whereas
the outer slope for the NFW profile is $r^{-3}$.
Increasing the FWHM of the beam from 5 to 8 arcminutes decreases the number of
objects.

Angular sizes of objects also depend on the gas distributions.  For small
angular sizes, both $\beta$ and NFW models predict a sharp cutoff which is due
to the intrinsic shape of the convolved $y$ profiles. The vast majority of 
objects have sizes below the beam FWHM and so are not resolved by {\it Planck}.
For the larger angular sizes, the $\beta$ model predicts more sources,
again due to the more extended distribution of gas in the outer regions of the
clusters.  Both $\tau$CDM and $\Lambda$CDM models predict similar peak numbers
of clusters but the $\tau$CDM model predicts more resolved clusters due to the
lower angular diameter distance.  

\section*{Acknowledgments}
We thank Gus Evrard for providing the lightcone halo catalogues (available at 
\verb+www.mpa-garching.mpg.de/Virgo/+), Domingos 
Barbosa and Antonio da Silva for many discussions, and Simon White for helpful 
comments.  The work presented in this paper was carried out as part of the 
programme of the Virgo Supercomputing Consortium
({\tt star-www.dur.ac.uk/$\sim$frazerp/virgo/}). STK is funded by PPARC and
PAT is a PPARC Lecturer Fellow.


\bsp
\end{document}